\documentclass[12pt,onecolumn,draftcls]{IEEEtran}
\usepackage{epsf,amssymb,amsfonts,cite,amsbsy}
\ifCLASSINFOpdf
\else
\fi
\usepackage{amsmath,amsthm}
\usepackage{epsf,amssymb,amsfonts}
\usepackage{graphicx}
\usepackage{cite,amsbsy}
\usepackage{latexsym}
\usepackage{verbatim}

\usepackage{algorithm}
\usepackage[noend]{algpseudocode}
\newtheorem{lemma}{Lemma}
\newtheorem{theorem}{Theorem}
\newtheorem{proposition}{Proposition}
\newcommand{\beq}{\begin{equation}}
\newcommand{\eeq}{\end{equation}}
\newcommand{\beqa}{\begin{eqnarray}}
\newcommand{\eeqa}{\end{eqnarray}}
\newcommand{\nn}{\nonumber}

\begin{document}
\title{\bf \LARGE A Study Of Optimal False Information Injection Attack On Dynamic State Estimation in Multi-Sensor Systems}
\author{Jingyang Lu and Ruixin Niu}
\maketitle

\begin{abstract}\label{abstract}
In this paper, the impact of false information injection is investigated for linear dynamic systems with multiple sensors. It is assumed that the system is unsuspecting the existence of false information and the adversary is trying to maximize the negative effect of the false information on Kalman filter's estimation performance. The false information attack under different conditions is mathematically characterized. For the adversary, many closed-form results for the optimal attack strategies that maximize Kalman filter's estimation error are theoretically derived. It is shown that by choosing the optimal correlation coefficients among the bias noises and allocating power optimally among sensors, the adversary could significantly increase Kalman filter's estimation errors. To be concrete, a target tracking system is used as an example in the paper. From the adversary's point of view, the best attack strategies are obtained under different scenarios, including a single-sensor system with both position and velocity measurements, and a multi-sensor system with position and velocity measurements. Under a constraint on the total power of the injected bias noises, the optimal solutions are solved from two perspectives: trace and determinant. Numerical results are also provided in order to illustrate the effectiveness of the proposed attack strategies. 
\end{abstract}
\IEEEpeerreviewmaketitle

\section{Introduction}
System state estimation in the presence of an adversary that injects false information into sensor readings has attracted much attention in wide application areas, such as target tracking with compromised sensors, secure monitoring of dynamic electric power systems and radar tracking and detection in the presence of jammers \cite{lu&niu:ciss2015}. This topic has been studied in \cite{liu&etal:ccs09, jia&etal:icassp11,kosut&etal_icsgc10,jia&etal_pesgm12,Rahman&Mohsenian-Rad_globecomm12,Jin&Lang:da2014,X&Song:TheMiMO2012,lu&niu:fusion2015}. In \cite{liu&etal:ccs09}, the problem of taking advantage of the power system configuration to introduce arbitrary bias to the system without being detected was investigated and inspired many researchers further study false information injection along this direction.  \cite{jia&etal:icassp11} shows the impact of malicious attacks on real-time electricity market concerning the locational marginal price and how the attackers can make profit by manipulating certain values of the measurements. Some certain strategies are also provied to find the optimal single attack vector. The relationship between the attackers and the control center was discussed in \cite{kosut&etal_icsgc10}, where both the adversary's attacking strategies and the control center's detection algorithms have been proposed. Refer to \cite{jia&etal_pesgm12} and  \cite{Rahman&Mohsenian-Rad_globecomm12} for more about false information attacks on the electricity market. Inspired by \cite{liu&etal:ccs09}, the data frame attack in which deleting the comprised sensors the defender system detects will make the system unobservable was formulated as a quadratically constrained quadratic program (QCQP) in \cite{Jin&Lang:da2014}. In \cite{X&Song:TheMiMO2012, lu&niu:fusion2015}, the relation between a target and a MIMO radar was characterized as a two-person zero-sum game.  However, in the aforementioned publications, only the problem of {\it static} system state estimation has been considered.

In this paper, for a linear {\em dynamic} system, we analyze the impact of the injected false information on Kalman filter's state estimation performance over time, which  has not got much attention in the literature. Some related publications exist on sensor management \cite{yang:op2012,yang:op2013,Mi&Claire:Sensor}, where the problem of arranging the sensors to minimize the covariance of the state estimation error so that a more accurate state estimate can be obtained is investigated. This problem is clearly opposite to the problem we study in the paper, where the goal for the adversary is to maximize the mean square state estimation error matrix, and to confuse Kalman filter. In \cite{lin&etal:aes06multisensor}, the problem of sensor bias estimation and compensation for target tracking has been addressed. Interested readers are referred to  \cite{lin&etal:aes06multisensor} and the references therein for details. 

In \cite{niu&huie_ssp12}, we have studied the impact of the injected biases on a Kalman filter's estimation performance, showing that if the false information is injected at a single time, its impact converges to zero as time goes on; if the false information is injected into the system continuously, the estimation error tends to reach a steady state. In \cite{lu&niu:fusion14}, we have found the best strategies for the adversary to attack Kalman filter system from the perspective of the trace of the mean squared error (MSE) matrix, and obtained some close-form results. Also in \cite{lu&niu:globalsip14}, based on the previous work, the problem is further refined regarding the determinant of MSE matrix, which the correlation among the elements is taken into consideration. In \cite{lu&niu:icasp2016,lu&niu:spie2016}, the Kalman filter system has been investigated regarding the system robustness in the case where sensor reading is continuously jammed by the false information using Greedy search and dynamic programming. However, the it is of great challenge to find the closed form solution in term of determinant of the MSE matrix and optimal solution to the case in which the Kalman filter system is compromised by the false information continuously. Considering the problems mentioned above, in this paper, our goal is to find the closed form optimal attack strategy for the adversary, which maximizes the impact of the false information injection on Kalman filter's state estimation from the determinant perspective. By adopting the objective function as the determinant of the MSE matrix, we change the problem significantly. As shown later in the paper, the optimal attack strategy  that maximizes the determinant of the MSE matrix is a function of Kalman filter's state estimation covariance and hence ''adaptive'' to Kalman filter; whereas that  maximizing the trace of the MSE matrix is not a function of Kalman filter's state estimation covariance. Previous works concentrated more on the situation where the adversary attacks the system by a single shot. In this paper, the problem of continuous attack is also investigated. 

The rest of paper is organized as follows. Section II generally describes the discrete-time linear dynamic system. Section III mathematically characterizes the impact of determined or random false information on Kalman filter's system. Section IV and V analyze how to get the best strategy to attack Kalman filter's system from trace and determinant cases standing from the perspective of the adversary. Under the constraint on the adversary's total sensor bias noise power, different strategies are proposed to maximize Kalman filter's mean squared state estimation error for different scenarios. Section VI provides the simulation results and Section VII concludes the paper.  
\section{Kalman filter System}

\section{System Model}
The discrete-time linear dynamic system can be described as below, 
\beq
	\mathbf{x}_{k+1} = \mathbf{F}_{k} \mathbf{x}_{k} + \mathbf{G}_{k} \mathbf{u}_{k} + \mathbf{v}_{k}
	\label{eq:gen_plant_eq}
\eeq
where ${\bf F}_{k}$ is the system state transition matrix, $\mathbf{x}_{k}$ is the system state vector at time $k$, $\mathbf{u}_{k}$ is a known input vector, $\mathbf{G}_{k}$ is the input gain matrix, and $\mathbf{v}_{k}$ is a zero-mean white Gaussian process noise with covariance matrix $E[\mathbf{v}_{k}\mathbf{v}_{k}^T] = \mathbf{Q}_{k}$. Let us assume that $M$ sensors are used by the linear system. The measurement at time $k$ collected by sensor $i$ is
\beq
	\mathbf{z}_{k,i} = \mathbf{H}_{k,i} \mathbf{x}_{k,i} + \mathbf{w}_{k,i} 
	\label{eq:meas_general}
\eeq
with $\mathbf{H}_{k,i}$ being the measurement matrix, and $\mathbf{w}_{k,i}$ a zero-mean white Gaussian measurement noise with covariance matrix $ E[\mathbf{w}_{k,i} \mathbf{w}_{k,i}^T] = \mathbf{R}_{k,i}$, for $i=1,\cdots, M$. We further assume that the measurement noises are independent across sensors. The matrices $\mathbf{F}_{k}$, $\mathbf{G}_{k}$, $\mathbf{H}_{k,i}$, $\mathbf{Q}_{k}$, and $\mathbf{R}_{k,i}$ are assumed to be known with proper dimensions.
In this paper, we assume that a bias $\mathbf{b}_{k,i}$ is injected by the adversary into the measurement of the $i$th  sensor at time $k$ intentionally. Therefore, the measurement equation (\ref{eq:meas_general}) becomes
\begin{eqnarray}
	\mathbf{z}'_{k,i} = \mathbf{H}_{k,i} \mathbf{x}_{k} + \mathbf{w}_{k,i} +\mathbf{b}_{k,i}=\mathbf{z}_{k,i}+\mathbf{b}_{k,i}  
	\label{eq:bias_model}
\end{eqnarray}
where $\mathbf{z}'_{k,i}$ is the corrupted measurement, $\mathbf{b}_{k,i}$ is either an unknown constant or a random variable independent of $\{\mathbf{v}_{k,i}\}$ and $\{\mathbf{w}_{k,i}\}$. 

For compactness, let us denote the system sensor observation as ${\bf z}_k=[{\bf z}_{k1}^T,\cdots, {\bf z}_{kM}^T]^T$, which contains the observations from all the $M$ sensors. Similarly, let us denote the system bias vector as ${\bf b}_k=[{\bf b}_{k1}^T,\cdots, {\bf b}_{kM}^T]^T$ which includes the biases at all the $M$ sensors. Correspondingly, the measurement matrix becomes
\beq
	{\bf H}_k=[{\bf H}_{k1}^T,\cdots,{\bf H}_{kM}^T]^T
	\label{eq:superhk}
\eeq 
With these notations, it is easy to convert (\ref{eq:meas_general}) and (\ref{eq:bias_model}) into the following equations respectively. 
\beq\label{eq:superbia}
	\mathbf{z}_{k} = \mathbf{H}_{k} \mathbf{x}_{k} + \mathbf{w}_{k} 
\eeq
and 
\beq
	\mathbf{z}'_{k} =\mathbf{z}_{k}+\mathbf{b}_{k}  
\eeq
Further, we have the measurement error covariance matrix corresponding to ${\bf w}_k$ is
\beq
	{\bf R}_k=\left[
	\begin{array}{lcr}
	 {\bf R}_{k,1}&\cdots&{\bf 0}  \\
	 \vdots&\ddots&	\vdots\\
	 {\bf 0}&\cdots& {\bf R}_{k,M}  \\
	 \end{array}
	\right]
	\label{eq:superrk}
\eeq
which is obtained by using the assumption that measurement noises are independent across sensors.

\section{Impact of False Information Injection}
\label{sec:general_sys}
In this paper, let us assume that the adversary attacks the system by injecting false information into the sensors while  unaware of such attacks. We start with the case where biases (${\bf b}_k$) are continuously injected into the system starting from a certain time $K$. Note that single injection is just a special case of continuous injection when ${\bf b}_k$ are set to be nonzero at time $K$ and zero otherwise.

In the continuous injection case, Kalman filter' extra state estimation error, which is caused by the continuous bias injection alone, is derived in \cite{niu:summer_ext_faculty_12} and provided as follows.  
\begin{proposition}
\label{pro:mse_bias}
Kalman filter's state estimation error at time $K+N$ is
\beqa 
\label{12}
\begin{split}
&{\hat{\mathbf{x}}'_{K+N|K+N}}-\mathbf{x}_{K+N}={\hat{\mathbf{x}}_{K+N|K+N}}-\mathbf{x}_{K+N}\\
&+\sum_{m=0}^N\left(\prod_{i=0}^{m-1}\mathbf{B}_{K+N-i}\right)\mathbf{W}_{K+N-m} \mathbf{b}_{K+N-m}\\
\label{eq:err}
\end{split}
\eeqa
where $\hat{\mathbf{x}}'_{K+N|K+N}$ is Kalman filter's state estimate in the presence of the bias sequence $\{{\bf b}_k\}$, $\hat{\mathbf{x}}_{K+N|K+N}$ is Kalman filter's state estimate in the absence of the bias, 
\beq
\mathbf{B}_{K}\triangleq \left(\mathbf{I}-\mathbf{W}_{K}\mathbf{H}_{K}\right) \mathbf{F}_{K-1},
\label{eq:A_K+N-i}
\eeq
 ${\bf I}$ is the identity matrix, and $\mathbf{W}_K$ is Kalman filter gain \cite{YBS:book} at time $K$. As a result, the extra state estimation error at time $K+N$ due to the continuous bias ${\bf b}_k$ injected at and after time $K$ is
\beq
\sum_{m=0}^N \left(\prod_{i=0}^{m-1} \mathbf{B}_{K+N-i}\right) \mathbf{W}_{K+N-m} \mathbf{b}_{K+N-m},  
\eeq
\end{proposition}

If \{${\bf b}_k$\} is a zero-mean, random, and independent sequence, the extra mean squared error (EMSE) at a particular time instant $K+N$ due to the bias alone is provided in the following proposition. Using the results from Proposition \ref{pro:mse_bias}, the proof of Proposition \ref{lem:mse_bias_continuous} is provided as well.

\begin{proposition}
\label{lem:mse_bias_continuous}
When the  bias sequence $\{{\bf b}_k\}$ is zero mean, random, and independent over time, the $EMSE$ at time $K+N$ due to the biases injected at and after time $K$, denoted as ${\bf A}_{K+N}$,  is
\beqa
	\mathbf{A}_{K+N}=\sum_{m=0}^{N}\mathbf{D}_{m}{\bf \Sigma}_{K+N-m} \mathbf{D}^T_{m}
	\label{random:xmse}
\eeqa
where 
\beqa
	\label{eq:dm}
	\mathbf{D}_{m}=\left(\prod_{i=0}^{m-1}\mathbf{B}_{K+N-i}\right)\mathbf{W}_{K+N-m}
\eeqa
 $\prod_{i=0}^{-1} \mathbf{B}_{K+N-i}={\bf I}$ is an identity matrix, and ${\bf \Sigma}_{K+N-m}$ is the covariance matrix of $\mathbf{b}_{K+N-m}$.
\end{proposition}
\noindent \underline{\em Proof Sketches:}
Let us denote $\tilde{\mathbf{x}}_{K+N|K+N}={\hat{\mathbf{x}}_{K+N|K+N}}-\mathbf{x}_{K+N}$ as Kalman filter's state estimation error in the absence of any false information, and 
\beq
	\mathbf{a}_{m}=\left(\prod_{i=0}^{m-1}\mathbf{B}_{K+N-i}\right)\mathbf{W}_{K+N-m}\mathbf{b}_{k+N-m}
\eeq
From (\ref{eq:err}), we can get
 \beqa
	\begin{split}
		& \mathbf{A}_{K+N}\\
		&=E\left[\left(\tilde{\mathbf{x}}_{K+N|K+N}+\sum_{m=0}^{N}\mathbf{a}_{m}\right)\left(\tilde{\mathbf{x}}_{K+N|K+N}+\sum_{n=0}^{N}\mathbf{a}_{n}\right)^T\right]\\&-E\left(\tilde{\mathbf{x}}_{K+N|K+N}\tilde{\mathbf{x}}_{K+N|K+N}^T\right)\\
		&=E\left(\tilde{\mathbf{x}}_{K+N|K+N}\sum_{n=0}^{N} \mathbf{a}_{n}^T\right)+E\left(\sum_{m=0}^{N} \mathbf{a}_{m} \tilde{\mathbf{x}}^T_{K+N|K+N}\right)\\
		&+E\left(\sum_{m=0}^{N}\sum_{n=0}^{N}\mathbf{a}_{m}\mathbf{a}_{n}^T\right)\\
		&=E\left(\sum_{m=0}^{N}\sum_{n=0}^{N} \mathbf{a}_{m}\mathbf{a}_{n}^T\right)
	\end{split}\nn
\eeqa
where the last line is due to the fact that ${\bf a}_m$ and ${\bf a}_n$ have zero mean, are independent from each other when $m\neq n$, and are independent from $\tilde{\mathbf{x}}_{K+N|K+N}$. Using this fact again, we further have 
\beqa
	E\left(\sum_{m=0}^{N}\sum_{n=0}^{N} \mathbf{a}_{m}\mathbf{a}_{n}^T\right)&=&E\left(\sum_{m=0}^{N} \mathbf{a}_{m}\mathbf{a}_{m}^T\right)\\
	&=&\sum_{m=0}^{N}\mathbf{D}_{m}{\bf \Sigma}_{K+N-m} \mathbf{D}^T_{m}\nn
\eeqa
where $\mathbf{D}_{m}$ has been defined in Proposition \ref{lem:mse_bias_continuous}.

\section{The Optimal Attack Strategy}
\subsubsection{Problem Formulation for a General Linear System \label{sec:general_opt}}
In this paper, we investigate the optimal attack strategy that an adversary can adopt to maximize the system estimator's estimation error. This problem can be formulated as a constrained optimization problem. Without loss of generality, let us consider that the attacker is interested in maximizing the  system state  estimation error at time $K$ right after a single false bias is injected at time $K$. In this case, we are interested in designing the injected random bias' covariance matrix such that 
\beqa 
	&&\max_{{\bf \Sigma}_K} \mathrm{Tr} \left[ {\bf P}_{K|K}+{\bf A}_K({\bf \Sigma}_K)\right]\nn\\ 
	&& s.t. \;\;\;\mathrm{Tr} ({\bf \Sigma}_K)=a^2
	\label{eq:max_trace}
\eeqa
where $a$ is a constant, $\mathrm{Tr}(\cdot)$ is the matrix trace operator, and ${\bf P}_{K|K}$ is Kalman filter's state estimation error covariance matrix at time $K$ in the absence of any false information. Note that it is meaningful to have a constraint on the trace of ${\bf \Sigma}_K$, since it can be deemed as the power of injected sensor bias ${\bf b}_K$, and a smaller power for ${\bf b}_K$ reduces the probability that the adversary is detected by the system estimator using an innovation based detector. Note that the optimization problem is equivalent to one that maximizes $\mathrm{Tr} \left( {\bf A}_K({\bf \Sigma}_K)\right)$, since ${\bf P}_{K|K}$ is not a function of ${\bf \Sigma}_K$, and trace is a linear operator. If one is more interested in the determinant of the estimation MSE matrix, a similar optimization problem can be easily formulated as follows. 
\beqa 
	&&\max_{{\bf \Sigma}_K} \left|  {\bf P}_{K|K}+{\bf A}_K({\bf \Sigma}_K)\right|\nn\\ 
	&& s.t. \;\;\;\mathrm{Tr} ({\bf \Sigma}_K)=a^2
	\label{eq:max_det}
\eeqa

\subsubsection{Equivalent Measurement in Multi-Sensor Systems\label{sec:equivalent}}
To simplify the mathematical analysis, it is helpful to derive the equivalent sensor measurement, which is a linear combination of the observations from all the sensors, and is a sufficient statistic containing all the information about the systems state. The equivalent sensor measurement vector and its corresponding covariance matrix should have much smaller dimension than the original measurement vector and its covariance, making the mathematical manipulation and derivation later in the paper much simpler.  In a information filter recursion \cite{YBS:book}, which is equivalent to Kalman filter recursion, we have  
\beqa
	\label{eq:information_filter}
	\hat {\bf y}_{k|k}=\hat {\bf y}_{k|k-1}+{\bf H}_k^T {\bf R}_k^{-1}{\bf z}_{k}
\eeqa
where $\hat {\bf y}_{k|k}={\bf P}_{k|k}^{-1} {\bf x}_{k|k}$ and $\hat {\bf y}_{k|k-1}={\bf P}_{k|k-1}^{-1} {\bf x}_{k|k-1}$. 
It is clear that $\hat{\bf y}_{k|k-1}$ represents the prior knowledge about the system state based on past sensor data, and the second term in (\ref{eq:information_filter}) represents the new information from the new sensor data ${\bf z}_k$, which can be expanded by using (\ref{eq:superhk}) and (\ref{eq:superrk}) as follows. 
\beqa
	\begin{split}
	&  {\bf H}^T_{k}{\bf R}_k^{-1}{\bf z}_{k}\\
	&=[{\bf H}_{k1}^T, \cdots,{\bf H}_{kM}^T]
	\left[
		\begin{array}{lcr}
			 {\bf R}_{k1}^{-1}&\cdots& {\bf 0}  \\
			 \vdots&\ddots&	\vdots\\
			 {\bf 0}&\cdots& {\bf R}_{kM}^{-1}
		 \end{array}
	\right] 
	\left[
		\begin{array}{lcr}
			{\bf z}_{k1}  	\\
					\vdots	\\
			{\bf z}_{kM}  	\\
		\end{array}
	\right] \\
	&=\sum_{i=1}^{M}{\bf H}^T_{ki}{\bf R}^{-1}_{ki} {\bf z}_{ki}
	\end{split}
\eeqa
In the following derivations, we skip the time index $k$ for simplicity. Our purpose is to find an equivalent measurement ${\bf z}_e$ such that 
\beq 
	{\bf z}_e={\bf H}_e {\bf x}+{\bf w}_e
\eeq
where ${\bf w}_e \sim {\cal N}({\bf 0}, {\bf R}_e)$, and 
\beqa
	\label{eq:fisher}
	{\bf H}^T_{e}{\bf R}^{-1}_{e} {\bf z}_{e}=\sum_{i=1}^{M}{\bf H}^T_{i}{\bf R}^{-1}_{i}{\bf z}_{i}
\eeqa
Let us consider two cases. First, suppose all the ${\bf H}_{i}$s are the same (${\bf H}_{i}={\bf H}$) , then it is natural to set ${\bf H}_e={\bf H}$. Note that a sufficient condition for (\ref{eq:fisher}) to be true is  
\beqa
	\label{eq:ze}
	{\bf z}_{e}={\bf R}_{e}\sum_{i=1}^{M} {\bf R}^{-1}_{i}{\bf z}_{i}
\eeqa
Taking the covariance on the both sides of (\ref{eq:ze}), we get
\beqa
	\begin{split}
	{\bf R}_e&={\bf R}_e \mathrm{cov}\left(\sum_{i=1}^{M}{\bf R}^{-1}_{i}{\bf z}_{i}\right){\bf R}^T_{e}\\
	&={\bf R}_e\left[\sum_{i=1}^{M}{\bf R}^{-1}_{i}{\bf R}_{i}({\bf R}^{-1}_{i})^{T}\right]{\bf R}^T_{e}
	\end{split}
\eeqa
This implies that 
\beqa
	\label{eq:rnew}
	{\bf R}_e=\left(\sum_{i=1}^{M}{\bf R}^{-1}_{i}\right)^{-1}
\eeqa
In the second case, let us assume that the system state ${\bf x}$ is observable based on the observations from all the sensors, meaning that the Fisher information matrix $\sum_{i=1}^{M} {\bf H}^T_{i}{\bf R}^{-1}_{i}{\bf H}_{i}$ is invertible. In this case, by setting ${\bf H}_e={\bf I}$, using (\ref{eq:fisher}), and following a similar procedure as in the first case, we have
\beqa
	\begin{split}
	\label{eq:measurement_equivalent}
	{\bf z}_{e}={\bf R}_e \sum_{i=1}^{M} {\bf H}_i^T {\bf R}^{-1}_{i} {\bf z}_{i}
	\end{split}
\eeqa
and 
\beqa
	{\bf R}_e=\left(\sum_{i=1}^{M} {\bf H}^T_{i}{\bf R}^{-1}_{i}{\bf H}_{i}\right)^{-1}
	\label{eq:re_total_obs}
\eeqa

\section{A Target Tracking Example }
In this paper, we give a concrete target tracking example. We assume that the target moves in a 1-dimensional space according to a discrete white noise acceleration model \cite{YBS:book}, which can still be described by the plant and measurement equations  given in (\ref{eq:gen_plant_eq}) and (\ref{eq:meas_general}). In such a system, the state is defined as $\mathbf{x_k}=[\xi_k \;\; \dot{\xi}_k]^T$, where $\xi_k$ and $\dot{\xi}_k$ denote the target's position and velocity at time $k$ respectively.  The input $\mathbf{u}_k$ is a zero sequence. 
The state transition matrix is 
\begin{eqnarray}
	\mathbf{F}=\left[\begin{array}{cc}
	1 &T\\
	0 &1 	
	\end{array}\right]
	\label{eq:F}
\end{eqnarray}
where $T$ is the time between measurements. The process noise is $\mathbf{v}_k=\mathbf{\Gamma} v_k$, where $v_k$ is a zero mean white acceleration noise, with variance $\sigma_{v}^{2}$, and the vector gain multiplying the scalar process noise is given by $ \mathbf{\Gamma}^T = \left[	T^{2}/2\;\;\;  	T \right]$.
The covariance matrix of the process noise is therefore $ \mathbf{Q}=\sigma^{2}_{v} \mathbf{\Gamma}  \mathbf{\Gamma}^{T}$.  

In this paper, we investigate the attack strategies for two scenarios. In the first scenario, only position measurements are available to the sensors, whereas in the second scenario, the sensors measure both position and velocity of the target.
\subsection{Attack Strageties Analysis From Trace perspective}

\subsubsection{Attack Strategy For Multiple Position Sensors\label{sec:max_trace}}
In this case, it is assumed that at each sensor, only the position measurement is available, so that $\mathbf{H}_i = \left[1 \;\; 0 \right]$. At each sensor, the measurement noise process is zero-mean, white, and with variance, $\sigma_{w_i}^{2}$.	In order to simplify the problem, we think of $\mathbf{z}_{e_k}$ as the equivalent measurement, which is a linear combination of the measurements from all  the sensors. Using the results we derived in Section \ref{sec:equivalent} for the first case, namely (\ref{eq:ze}) and (\ref{eq:rnew}), the measurement equation (\ref{eq:bias_model}) becomes
\beq
	{z}'_{k}={z}_{ek}+{b}_{ek}
	\label{eq:equivalent_measurement}
\eeq
where 
\beq
	z_{ek}=\sum_{m=0}^M c_i {z}_{ki}\ 
	\label{eq:euivalent_sensor_z}
\eeq
\beq
	{b}_{ek}=\sum_{m=0}^M c_{i} {b}_{ki}\
	\label{eq:euivalent_sensor_b}
\eeq
and 
\beq
	c_{i}=\dfrac{1/{\sigma^2_{w_i}}}{\sum_{j=1}^M  \left(1/{\sigma^2_{w_j}}\right)}
	\label{eq:coefficient}
\eeq
which is the corresponding coefficient/weight for the $i$th sensor. 
In this target tracking problem, let us first consider the strategy that maximizes the trace of the Kalamn filter estimation error, which is the solution of (\ref{eq:max_trace}) in Section \ref{sec:general_opt}. In this case, 
\beq 
	{\bf \Sigma}_K=\left[\begin{array}{cccc}
	\sigma_{b_1}^2& \rho_{12}\sigma_{b_1}\sigma_{b_2}&\cdots& \rho_{1M}\sigma_{b_1}\sigma_{b_M}\\
	\rho_{12}\sigma_{b_1}\sigma_{b_2}& \sigma_{b_2}^2&\cdots& \rho_{2M}\sigma_{b_2}\sigma_{b_M}\\
	\vdots& \vdots& \ddots & \vdots\\
	\rho_{1M}\sigma_{b_1}\sigma_{b_M}& \rho_{2M}\sigma_{b_2}\sigma_{b_M}&\cdots& \sigma_{b_M}^2
	\end{array}
	 \right]
\eeq
where $\sigma_{b_i}^2$ is the variance of the random bias injected at the $i$th sensor ($b_i$), and $\rho_{ij}$ is the correlation coefficient between $b_i$ and $b_j$. 
Therefore, (\ref{eq:max_trace}) is equivalent to  
\beqa 
	&&\max \mathrm{Tr} \left[ {\bf A}_K \right]\nn\\ 
	&& s.t. \;\;\;\sum_{i=1}^{M} \sigma_{b_i}^2 = a^2\nn\\
	&&\;\;\;\ -1\leq \rho_{ij} \leq 1, \;\;\mathrm{for} \;\; 1\leq i, j\leq M
	\label{eq:max_trace_ak}
\eeqa
To simplify this problem, we first use the equivalent measurement to convert the multi-sensor problem to a single sensor problem. Namely, in Proposition \ref{lem:mse_bias_continuous} by replacing 
\[{\bf H}_k=\left[\begin{array}{cc}
1&0\\
\vdots&\vdots\\
1&0
\end{array}
\right]\]
with ${\bf H}_e=[1\;0]$, and replacing ${\bf \Sigma}_K$ with 
\beqa
	\label{eq:sigmaek}
	{ \Sigma}_{e_K}&=& E[b_{e_K}^2] \\
	&=&E\left[\left(\sum_{i=1}^M c_{i}b_{i}\right)^2\right]\nn\\
	&=& \sum_{i=1}^M c_{i}^2 \sigma_{b_{i}}^2+ \sum_{i} \sum_{j\neq i} 2\rho_{ij}c_{i}c_{j}\sigma_{b_{i}}\sigma_{b_{j}}\nn
\eeqa
we can easily show that ${\bf A}_K={\bf D}_0 \Sigma_{e_K} {\bf D}_0^T$.  Since ${ \Sigma}_{e_K}$ is a scalar and ${\bf D}_0$ is not a function of ${\bf \Sigma}_K$, maximizing the trace of ${\bf A}_K$ is equivalent to maximizing ${ \Sigma}_{e_K}$. 

\begin{figure}[!t]
	\centering
	 \includegraphics[width=2.5in]{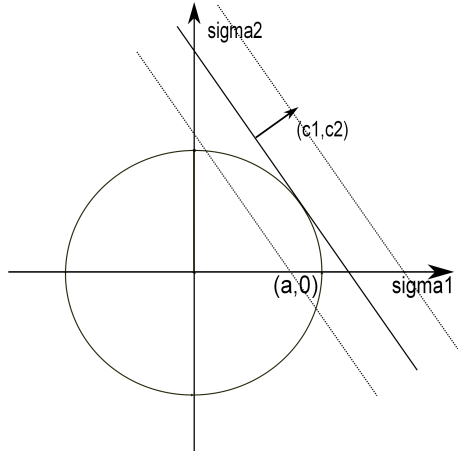}
	\caption{Geometric solution for systems with two sensors.}
	\label{fig:2_dimension}
\end{figure}

First, let us consider the case where the random biases at different sensors are independent,  meaning that $\rho_{i,j}=0$ for $1\leq i,j \leq M$. The optimal strategy for the adversary in this case is clearly to put all the bias power to the sensor with the largest coefficient $c_{i}$: 
\begin{proposition}
For a system with $M$ sensors, if the adversary injects independent random noises, the best strategy  is to allocate all the power to the sensor with smallest noise variance.
\end{proposition}
Next, let us consider the more general case where the random biases are dependent. By inspecting (\ref{eq:sigmaek}), it is clear that to maximize ${ \Sigma}_{e_K}$, we need to set all the $\rho_{ij}$s to 1. As a result, (\ref{eq:sigmaek}) becomes
\beqa
	{ \Sigma}_{e_K}=\left(\sum_{i=1}^M c_{i} \sigma_{b_i}\right)^2
	\eeqa
Now, the optimization problem in (\ref{eq:max_trace_ak}) has been converted to the following problem: 
\beqa 
	&&\max \left(\sum_{i=1}^M c_{i} \sigma_{b_i}\right)^2 \nn\\ 
	&& s.t. \;\;\;\sum_{i=1}^{M} \sigma_{b_i}^2 = a^2
\eeqa
The above problem can be solved by using standard constrained optimization techniques \cite{boyd&Vandenberghe:book} based on gradient and Hessian, which are rather involved. Here we solve the problem using a much simpler geometric solution, which has been shown to give the same solution as that by the standard optimization techniques.  We start with the simplest case with two sensors, in which we need to solve the following optimization problem.
\begin{eqnarray}
	\max & 
	{c}_{1}\sigma_{b_1}+{c}_{2}\sigma_{b_2}\\
	\text{s.t.} & 
	 \begin{aligned}
	  \sigma_{b_1}^2+\sigma_{b_2}^2=a^2\\
	  \end{aligned}\nn
\end{eqnarray}
We can get the optimal solution by analyzing the problem geometrically with the norm vector $(c_{1},c_{2})^T$ of the objective function as shown in the Fig. \ref{fig:2_dimension}. The constraint of the problem is represented by the circle with a radius of $a$. We move the line $l_{1}$ with the slope  $-\dfrac{c_{1}}{c_{2}}$ to get the largest intercept between $l_{1}$ and $\sigma_{2}$ axis under the constraint that there is an intersection between the circle and the line $l_{1}$. The corresponding optimal solution is found when $l_{1}$ becomes a tangent line to the circle, which is  
\beq
	\begin{split}
	\sigma_{1}=\frac{c_{1}a}{\sqrt{c^2_{1}+c^2_{2}}}\\
	\sigma_{2}=\frac{c_{2}a}{\sqrt{c^2_{1}+c^2_{2}}}\\
	\end{split}
	\label{op_solu_2}
\eeq

For a system with arbitrary number of sensors, we can repeat the same procedure to find the optimal solution by using hyperplanes and hyperspheres. In general, the optimal attack strategy can be found and summarized as follows. 
\begin{theorem}
For a system with $M$ sensors, the optimal strategy for the adversary is to inject dependent random noises with a pairwise correlation coefficient of $1$. The random bias power is allocated such that
\beq
	\sigma_{b_i}=\frac{c_{i}a}{\sqrt{ \sum_{j=1}^M c^2_{j} }}, \;\;\; \mathrm{for}\; i=1,\cdots,M.
\eeq
\end{theorem}
\subsubsection{Attack Strategy For A Single Position And Velocity Sensor}
In this case, let us assume that the sensors collect both position and velocity measurements of the target. Therefore, the measurement matrix for the $i$th sensor is ${\bf H}_i={\bf I}_2$, where ${\bf I}_2$ is a $2\times 2$ identity matrix. At the $i$th sensor, the  adversary injects  the bias noise vector $\mathbf{b}_{k_i}$ to the sensor measurement ${\bf z}_{k_i}$, where $\mathbf{b}_{k_i}=[b_{p_i} \;\;b_{v_i}]^T$ consists biases in position and velocity measurements. Let us assume that the system bias vector  ${\bf b}_k=[{\bf b}_{k1}^T,\cdots, {\bf b}_{kM}^T]^T$ is zero-mean and has a $2M\times 2M$ covariance matrix ${\bf \Sigma}_K$. Further, the $(i,\;j)$th $2\times 2$ submatrix for ${\bf \Sigma}_K$ is defined as 
\beq 
	{\bf \Sigma}_K(i,j)=\left[\begin{array}{cccc}
	\rho_{{b_p}_i,{b_p}_j}\sigma_{b_{p_i}}\sigma_{b_{p_j}}& \rho_{{b_p}_i,{b_v}_j}\sigma_{{b_p}_i}\sigma_{{b_v}_j}\\
	\rho_{{b_v}_i,{b_p}_j}\sigma_{{b_v}_i}\sigma_{b_{p_j}}& \rho_{{b_v}_i,{b_v}_j}\sigma_{{b_v}_i}\sigma_{b_{v_j}}\\
	\end{array}
	 \right]
	\label{eq:sub_matrix_sigma}
\eeq
for $1\leq i,\;j\leq M$. $\sigma_{{b_p}_i}$ and $\sigma_{{b_v}_i}$ are the position and velocity bias noise standard deviations at the $i$th sensor respectively. The $\rho$s are defined as the proper correlation coefficients between components of the bias vector, and $\rho_{{b_p}_i,{b_p}_i}=\rho_{{b_v}_i,{b_v}_i}=1$, for $1\leq i \leq M$. Since the position bias $b_p$  and velocity bias $b_v$ have different units, we need an appropriate constraint for bias noise power. Here we assume that the total noise power is defined as 
\beq
 \sum_{i=1}^{M} \sigma_{b_{p_i}}^2+ T^2 \sigma_{b_{v_i}}^2
\label{eq:total_power_pv}
\eeq 
Note that this is a meaningful power definition, since the two terms in the above equation has the same unit. Recall that according to the target tracking system plant equation and ignoring the system process noise, we have $\xi_{k+1}=\xi_k+T \dot{\xi}_k$. Therefore, the power defined in (\ref{eq:total_power_pv}) can be interpreted as the summation of the extra mean squared errors for the position estimate caused by independent bias injections. We can see that the best attack strategy derived under a constraint on power defined in (\ref{eq:total_power_pv}) can be easily adjusted and extended for other power definitions, as long as in the new definition, the second term is proportional to $T^2 \sigma_{b_{v_i}}^2$. 

As we can use the equivalent sensor to represent the multiple sensors, we focus on the single-sensor case first. If we are interested in the case of $N=0$, maximizing the trace of ${\bf A}_K$ is equivalent to maximize the $ \mathbf{W}_{K}{\bf \Sigma}_{K}\mathbf{W}^{T}_{K}$. We assume that  the adversary knows the system models and the prior information ${\bf P}_{0|0}$ at  time zero, so that he/she can calculate the offline Kalman filter gain matrix ${\bf W}_k$ recursively. Therefore, the best strategy the adversary can adopt to attack the system is the  solution to the following optimization problem: 
\beqa 
&&\max_{{\bf \Sigma}_{K}} \mathrm{Tr} \left[ \mathbf{W}_{K}{\bf \Sigma}_{K}\mathbf{W}^{T}_{K} \right]\nn\\ 
&& s.t. \;\;\; \sigma_{b_p}^2+T^2\sigma_{b_v}^2 = a^2\nn\\
&&\;\;\;\;\;\ -1\leq \rho_{b_p,b_v} \leq 1\nn\\
&&\;\;\;\;\;\ \sigma_{b_p},\sigma_{b_v} > 0 
\label{eq:max_trace_vec_single}
\eeqa   
where 
\beq 
\boldsymbol{\Sigma}_K=\left[\begin{array}{cc}
\sigma_{b_p}^2& \rho_{b_p,b_v}\sigma_{b_p}\sigma_{b_v}\\
\rho_{b_p,b_v}\sigma_{b_p}\sigma_{b_v}& \sigma_{b_v}^2\\
\end{array}\right]
\eeq
and
\beq
\mathbf{W}_K=\left[\begin{array}{cc}
w_{11}& w_{12}\\
w_{21}& w_{22}\\
\end{array}
 \right]
\eeq
It is easy to show that 
\beqa
\begin{split}
\mathrm{Tr}& \left[\mathbf{W}_{K}{\bf \Sigma}_{K}\mathbf{W}^{T}_{K} \right]=\mathrm{Tr} \left[\mathbf{W}^{T}_{K} \mathbf{W}_{K}{\bf \Sigma}_{K} \right]\\
&=(w_{11}^2+w_{21}^2)\sigma_{b_p}^2+(w_{12}^2+w_{22}^2)\sigma_{b_v}^2\\
&+2(w_{11}w_{12}+w_{21}w_{22})\rho_{b_p,b_v}\sigma_{b_p}\sigma_{b_v}\\
\end{split}
\eeqa
According to the sign of  $(w_{11}w_{12}+w_{21}w_{22})$, we can set the value of the $\rho_{b_p, b_v}$ to maximize the objective function. For example, if $(w_{11}w_{12}+w_{21}w_{22})$ is positive, we set $\rho_{b_p, b_v}=1$ and the optimization problem becomes
\beqa 
&&\max  (w_{11}\sigma_{b_p}+w_{12}\sigma_{b_v})^2+(w_{21}\sigma_{b_p}+w_{22}\sigma_{b_v})^2\nn\\ 
&& s.t. \;\;\; \sigma_{b_p}^2+T^2\sigma_{b_v}^2 = a^2\\
&& \;\;\; \sigma_{b_p},\sigma_{b_v} \geq 0 \nn
\label{eq:max_trace_vec_simplify}
\eeqa

To solve this constrained optimization problem, let us first denote
\beqa
	\begin{split}
		w_{11}^2+w_{21}^2=\beta_{1}\\
		w_{12}^2+w_{22}^2=\beta_{2}\\
		w_{11}w_{12}=\alpha_{1}\\
		w_{21}w_{22}=\alpha_{2}\\
	\end{split}
\eeqa   
The constraint in (\ref{eq:max_trace_vec_single})  can be written as 
\beqa
	\frac{\sigma_{b_p}^2}{T^2}+\sigma_{b_v}^2=\frac{a^2}{T^2}=a_{1}^2
\eeqa
Now we set $\sigma_{b_p}=a_{1}T\sin(\theta)$ and $\sigma_{b_v}=a_{1}\cos(\theta)$. Plugging $\sigma_{b_p}$ and $\sigma_{b_v}$ into the objective function, we have the following equivalent optimization problem 
\beqa 
	&&\max_\theta   a_{1}^2\left[\frac{\beta_{1}T_{1}^2+\beta_{2}}{2}+A\sin(2\theta+\phi) \right]\nn\\ 
	&& s.t. \;\;\; 0\leq\theta\leq \frac{\pi}{2}
	\label{eq:max_trace_vec_simplify}
\eeqa
where
\beqa
	 A&=&\sqrt{\frac{1}{4}\left(\beta_{2}-\beta_{1}T^2\right)^2+T^2(\alpha_{1}+\alpha_{2})^2}\\
	\tan(\phi)&=&\frac{\beta_{2}-\beta_{1}T^2}{2T(\alpha_{1}+\alpha_{2})}
\eeqa
Clearly, the optimal solution is 
\beq 
	\theta^*=\frac{\pi}{4}-\frac{\phi}{2}
\eeq

We summarize this result in the following theorem.
\begin{theorem}
\label{the:single_vector}
For a system with one sensor observing position and velocity of the target, the optimal strategy for the adversary is to inject random noise that has dependent position and velocity components. If $w_{11}w_{12}+w_{21}w_{22}>0$,  the correlation coefficient $\rho_{b_{p},b_v}$ should be set as $1$, and the random bias power is allocated such that
\begin{eqnarray}
	\label{eq:optimal_pv_single}
	&& \sigma_{b_p}=a\sin(\theta^*) \\
	&& \sigma_{b_v}=\frac{a}{T}\cos(\theta^*)\nn\\
	&& \theta^*=\frac{\pi}{4}-\frac{\phi}{2}\nn\\
	&&\phi=\arctan\left[\frac{\beta_{2}-\beta_{1}T^2}{2T(\alpha_{1}+\alpha_{2})}\right]\nn\\
	&& w_{11}^2+w_{21}^2=\beta_{1}\nn\\
	&& w_{12}^2+w_{22}^2=\beta_{2}\nn\\
	&& w_{11}w_{12}=\alpha_{1}\nn\\
	&& w_{21}w_{22}=\alpha_{2}\nn
\end{eqnarray}
When $w_{11}w_{12}+w_{21}w_{22}<0$, we should set $\rho_{b_p,b_v}=-1$ and set $\alpha_{1}=-w_{11}w_{12}$ and $\alpha_{2}=-w_{21}w_{22}$. The rest of the equations in formula (\ref{eq:optimal_pv_single}) remains the same.
\end{theorem}
\subsubsection{Attack Strategy For Multiple Position And Velocity Senors}
In this case, $M=2$, and the measurement matrix is ${\bf H}=[{\bf I}_2\; {\bf I}_2]^T$. The measurement covariance matrix for the $i$th sensor is assumed to be 
\beq
	{\bf R}_{i}=\left[\begin{array}{ccc}
	\sigma^{2}_{p_i}& 0\\
	0& \sigma^{2}_{v_i}\\
	\end{array}\right]
\eeq
Now, according to (\ref{eq:re_total_obs}), we have 
\beq
	\begin{split}
		{\bf R}_{e}&=[{\bf R}^{-1}_{1}+{\bf R}^{-1}_{2}]^{-1}\\
		&=\left[\begin{array}{cc}
		\left(\sigma^{-2}_{p_1}+\sigma^{-2}_{p_2}\right)^{-1}& 0\\
		0& \left(\sigma^{-2}_{v_1}+\sigma^{-2}_{v_2}\right)^{-1}\\
		\end{array}\right]
	\end{split}
\eeq
According to (\ref{eq:measurement_equivalent}), we define 
\beq
	\begin{split}
		{\bf C}_{i}&={\bf R}_{e}{\bf H}_i^T {\bf R}^{-1}_{i}\\
		&=\left[\begin{array}{ccc}
		\frac{\sigma^{-2}_{p_i}}{\sigma^{-2}_{p_1}+\sigma^{-2}_{p_2}}& 0\\
		0&\frac{\sigma^{-2}_{v_i}}{\sigma^{-2}_{v_1}+\sigma^{-2}_{v_2}}\\
		\end{array}\right]
	\end{split}
\eeq
as the weighting matrix for the $i$th sensor's observation ${\bf z}_i$.
Further, we define  
\beq
	\label{eq:v_coe}
	\begin{split}
		c_{p_i}={\bf C}_{i}(1,1)\\
		c_{v_i}={\bf C}_{i}(2,2)
	\end{split}
\eeq
both of which are positive numbers. The equivalent noise injection is therefore
\beq 
	{\bf b}_{eK}=\sum_{i=1}^{2} {\bf C}_i {\bf b}_{K_i}
\eeq
So the covariance matrix of the equivalent bias vector is
\beq
	{\bf \Sigma}_{eK}=\sum_{i=1}^{2} \sum_{i=j}^{2} {\bf C}_i  {\bf \Sigma}_K(i,j) {\bf C}_j^T
\eeq
where ${\bf \Sigma}_K(i,j)$ has been defined in (\ref{eq:sub_matrix_sigma}). It can be shown that
\beq 
	{\bf \Sigma}_{eK}=\left[\begin{array}{cc}
	s_1&s_2 \\
	s_2& s_3\\
	\end{array}
	 \right]
\eeq 
Where
\beq
	\begin{split}
		s_1=c_{p_1}^{2}\sigma_{b_{p_1}}^{2}+c_{p_2}^2\sigma_{b_{p_2}}^2+2\rho_{b_{p_1},b_{p_2}}c_{p_1}c_{p_2}\sigma_{b_{p_1}}\sigma_{b_{p_2}}\\
		s_3=c_{v_1}^2\sigma_{b_{v_1}}^2+c_{v_2}^2\sigma_{b_{v_2}}^2+2\rho_{b_{v_1},b_{v_2}}c_{v_1}c_{v_2}\sigma_{b_{v_1}}\sigma_{b_{v_2}}
	\end{split}
\eeq
\beq
	\begin{split}
		s_2&=c_{p_1}c_{v_1}\rho_{b_{p_1},b_{v_1}}\sigma_{b_{p_1}}\sigma_{b_{v_1}}+c_{p_1}c_{v_2}\rho_{b_{p_1},b_{v_2}}\sigma_{b_{p_1}}\sigma_{b_{v_2}} \\ &+c_{p_2}c_{v_1}\rho_{b_{p_2},b_{v_1}}\sigma_{b_{p_2}}\sigma_{b_{v_1}}+c_{p_2}c_{v_2}\rho_{b_{p_2},b_{v_2}}\sigma_{b_{p_2}}\sigma_{b_{v_2}}
	\end{split}
\eeq
The optimization problem can be written as follows.
\beqa
	\label{eq:max_trace_vec}
	&&\max_{{\bf \Sigma}_{eK}} \mathrm{Tr} \left[ \mathbf{W}_{eK}{\bf \Sigma}_{eK}\mathbf{W}^{T}_{eK} \right]\\ 
	&& s.t. \;\;\; \sigma_{b_{p_1}}^2+\sigma_{b_{p_2}}^2+T^2\sigma_{b_{v_1}}^2+T^2\sigma_{b_{v_2}}^2 = a^2,\nn\\
	&&\;\;\;\;\;\ -1\leq \rho_{p_i,v_j} \leq 1, \nn\\
	&&\;\;\;\;\;\ -1\leq \rho_{v_i,v_j} \leq 1, \nn\\
	&&\;\;\;\;\;\ -1\leq \rho_{p_i,p_j} \leq 1, \nn\\
	&&\;\;\;\;\;\ \sigma_{p_i},\sigma_{v_i}\geq 0, \;\;\;\forall i,j \in \{1,2\}\nn
\eeqa  
where 
\beq 
	\mathbf{W}_{eK}=\left[\begin{array}{cc}
	w_{11}& w_{12}\\
	w_{21}& w_{22}\\
	\end{array}
	 \right]
\eeq
is Kalman filter gain calculated using the equivalent measurement covariance matrix ${\bf R}_e$ and equivalent measurement matrix ${\bf H}_e$. It is easy to show that
\beqa
	\label{eq:obj_vector}
	&&\mathrm{Tr}\left[\mathbf{W}_{K}{\bf \Sigma}_{K}\mathbf{W}^{T}_{K} \right]=\mathrm{Tr} \left[\mathbf{W}^{T}_{K} \mathbf{W}_{K}{\bf \Sigma}_{K} \right]\\
	&&=(w^2_{11}+w^2_{21})^2s_1+(w^2_{12}+w^2_{22})^2s_3 \nn\\
	&&+2(w_{11}w_{12}+w_{21}w_{22})s_2 \nn
\eeqa
Clearly, all the $\rho$s that appear in $s_1$ and $s_3$ should be set as 1 to maximize the objective function. The optimal values for $\rho$s in $s_2$ depend on Kalman filter gain ${\bf W}_{eK}$. More specifically, when $w_{11}w_{12}+w_{21}w_{22}>0$, all the $\rho$s that appear in  $s_2$ should be set to  $1$; otherwise, they should be set as $-1$. Let us first suppose that $w_{11}w_{12}+w_{21}w_{22}>0$ is true, then we have 
\beqa
	\begin{split}
	&\mathrm{Tr}\left[\mathbf{W}_{K}{\bf \Sigma}_{K}\mathbf{W}^{T}_{K} \right]=(w^2_{11}+w^2_{21})^2(c_{p_1}\sigma_{p_1}+c_{p_2}\sigma_{p_2})^2\\
	&+(w^2_{12}+w^2_{22})^2(c_{v_1}\sigma_{v_1}+c_{v_2}\sigma_{v_2})^2\\
	&+2(w_{11}w_{12}+w_{21}w_{22})(c_{p_1}c_{v_1}\sigma_{p_1}\sigma_{v_1}+c_{p_1}c_{v_2}\sigma_{p_1}\sigma_{v_2}\\
	&+c_{p_2}c_{v_1}\sigma_{p_2}\sigma_{v_1}+c_{p_2}c_{v_2}\sigma_{p_2}\sigma_{v_2})
	\end{split}
\eeqa

So far, we have converted the objective function in (\ref{eq:max_trace_vec}), which involves 10 variables to one that involves only 4 variables. Considering that the power constraint reduces one degree of freedom, we only need to solve an optimization problem in a 3-dimensional space. 
\subsubsection{Strategy For A Single Sensor With Multiple Time Attack}
Based on Proposition \ref{lem:mse_bias_continuous}, we get the extra mean square matrix,
\begin{eqnarray}
	\mathbf{A}_{K+N}=\sum_{m=0}^{N}\mathbf{D}_{m}{\bf \Sigma}_{K+N-m} \mathbf{D}^T_{m} \nn
\end{eqnarray}
Suppose at the time $K$, the adversary wants to attack the system continuously from time $K$ to $K+N$, the weight for different time is $\alpha_{i}, i \in N$, as shown below,
\begin{eqnarray}
	&&\mathbf{A}^{'}_{K+0}=\alpha_{0}(\mathbf{D}_{0}{\bf \Sigma}_{K} \mathbf{D}^T_{0})\nn\\
	&&\mathbf{A}^{'}_{K+1}=\alpha_{1}(\mathbf{D}_{0}{\bf \Sigma}_{K+1} \mathbf{D}^T_{0}+\mathbf{D}_{1}{\bf \Sigma}_{K} \mathbf{D}^T_{1})\\
	&&...\nn\\
	&&\mathbf{A}^{'}_{K+N}=\alpha_{N}(\mathbf{D}_{0}{\bf \Sigma}_{K+N} \mathbf{D}^T_{0}+...+\mathbf{D}_{N}{\bf \Sigma}_{K} \mathbf{D}^T_{N})\nn
\end{eqnarray}
where $\sum^{N}_{m=0}\alpha_{m}=1$. So the objective function in the multi-shot attack case is the trace of the weighted sum of the EMSE matrices at different time points that is $\sum_{m=0}^{N}\alpha_{m}\mathbf{A}_{K+m}=\sum_{m=0}^{N}\mathbf{A}^{'}_{K+m}$. It is equivalent to maximize the trace of the weighted sum of the MSE matrices of the state estimates, because once the system reaches its steady state, ${\bf P}_{K+m|K+m}$ becomes constant, and the weighted sum of ${\bf P}_{K+m|K+m}$ will remain the same. First we study the case where the system has position sensors which are being attacked, so all the items above are scalars. Using lower case $d,\sigma_p^2$ to denote $\mathbf{D},{\bf \Sigma}$, we can formulate the optimization problem below,
\begin{eqnarray}
\max_{\sigma_{p_K}, \cdots, \sigma_{p_{K+N}} }	&&\sum_{m=0}^{N}\alpha_{m}\mathbf{A}_{K+m}=\sum_{m=0}^{N}\mathbf{A}^{'}_{K+m}\\
	&&=\sigma^2_{p_{K}}(\alpha_{0}d^2_{0}+\alpha_{1}d^2_{1}+...+\alpha_{N}d^2_{N})\nn\\
	&&+\sigma^2_{p_{K+1}}(\alpha_{1}d^2_{0}+\alpha_{2}d^2_{1}+...+\alpha_{N}d^2_{N-1})\nn\\
	&&+\sigma^2_{p_{K+2}}(\alpha_{2}d^2_{0}+\alpha_{3}d^2_{1}+...+\alpha_{N}d^2_{N-2})\nn\\
	&&+...\nn\\
	&&+\sigma^2_{p_{K+N}}(\alpha_{N}d^2_{0})\nn\\
	s.t.&& \sum_{m=K}^{K+N}\sigma^2_{p_{m}}\le a^2\nn\\
	&&\sum^{N}_{m=0}\alpha_{m}=1 \nn
\end{eqnarray}

The adversary can allocate the power based on the coefficients of the variance variables at different time. For example, if the weights $\alpha_{m}'s$ are all the same, the best strategy is to allocate all the power to the sensors at the first beginning (at time K) because the coefficient for $\sigma^2_{p_{K}}$ is the largest. Second, if the sensors measure both position and velocity, and the attacker aims to attack the system with position and velocity false information, the optimization problem can be characterized as below,

\begin{eqnarray}
\max_{\boldsymbol{\Sigma}_K, \cdots, \boldsymbol{\Sigma}_{K+N}}	&& \mathrm{Tr} \left[ \sum_{m=0}^{N}\alpha_{m}\mathbf{A}_{K+m}\right]= \mathrm{Tr}\left[\sum_{m=0}^{N}\mathbf{A}^{'}_{K+m}\right]\\
	&&= \mathrm{Tr}\left[{\bf \Sigma}_{{K}}(\alpha_{0}\mathbf{D}^T_{0}\mathbf{D}_{0}+...+\alpha_{N}\mathbf{D}^T_{N}\mathbf{D}_{N})\right]\nn\\
	&&+\mathrm{Tr}\left[{\bf \Sigma}_{{K+1}}(\alpha_{1}\mathbf{D}^T_{0}\mathbf{D}_{0}+...+\alpha_{N}\mathbf{D}^T_{N-1}\mathbf{D}_{N-1})\right]\nn\\
	&&+\mathrm{Tr}\left[{\bf \Sigma}_{{K+2}}(\alpha_{2}\mathbf{D}^T_{0}\mathbf{D}_{0}+...+\alpha_{N}\mathbf{D}^T_{N-2}\mathbf{D}_{N-2})\right]\nn\\
	&&+...\nn\\
	&&+\mathrm{Tr}\left[{\bf \Sigma}_{{K+N}}(\alpha_{N}\mathbf{D}^T_{0}\mathbf{D}_{0})\right]\nn\\
	s.t.&& \sum_{m=K}^{K+N}\sigma^2_{p_{m}}+T^2\sigma^2_{v_{m}}\le a^2\nn\\
	&&\sum^{N}_{m=0}\alpha_{m}=1 \nn
\end{eqnarray}
where ${\bf \Sigma}_{m}$ and $\mathbf{D}^T_{j}\mathbf{D}_{j}$ are positive semidefinite matrices, so $\mathrm{Tr}\left[{\bf \Sigma}_{{m}}(\mathbf{D}^T_{j}\mathbf{D}_{j})\right]\ge 0$ all the time. The trace function $\mathrm{Tr(\cdot)}$ is a monotonically increasing function of the positive semidefinite matrix. So the best strategy for the adversary to attack the system is to put all the power at the time with the largest positive semidefinite matrix.

\subsection{Attack Strategies from Determinant Perspective}
  
\subsubsection{Attack Strategy For Multiple Position Sensors}  
We are also interested in the effect of bias information on Kalman filter's estimation MSE from the determinant perspective. By using the equivalent measurement approach as in Section \ref{sec:max_trace}, we have 
\beqa
	\begin{split}
	\label{eq:Deter}
	&|{\bf P}_{K|K} + {\bf A}_K|=|{\bf P}_{K|K}+{\Sigma}_{eK}\mathbf{D}_{0}\mathbf{D}^T_0|\\
	&=|{\bf P}_{K|K}||{\bf I}+{\Sigma}_{eK}\mathbf{D}_{0}{\bf P}^{-1}_{K|K}\mathbf{D}_0^T|\\
	\end{split}
\eeqa
where $\mathbf{D}_{0}$ is defined in Proposition \ref{lem:mse_bia}. ${\Sigma}_{eK}$ is defined in (\ref{eq:sigmaek}). As ${\bf P}_{K|K}$ is constant and positive definite, $\mathbf{D}_{0}{\bf P}^{-1}_{K|K}\mathbf{D}_0^T$ is  positive semidefinite meaning that all the eigenvalues of the $\mathbf{D}_{0}{\bf P}^{-1}_{K|K}\mathbf{D}_0^T$ are non-negative. First, let us
denote ${\bf C}$ as a square matrix whose columns are the eigenvectors of 
$\mathbf{D}_{0}{\bf P}^{-1}_{K|K}\mathbf{D}_0^T$. Then through eigendecomposition, (\ref{eq:Deter}) can be written  concisely as,
\beqa
\label{eq:Deter_concise}
	\begin{split}
		&|{\bf P}_{K|K}||{\bf C}{\bf I}{\bf C}^{-1}+{\Sigma}_{eK}{\bf C}{\bf \Lambda} {\bf C}^{-1}|\\
		&=|{\bf P}_{K|K}||{\bf I}+{\Sigma}_{eK} \mathbf{\Lambda}|\\
	\end{split}
\eeqa
where $\mathbf{\Lambda}$ is a diagonal matrix whose  diagonal elements are the eigenvalues of the $\mathbf{D}_{0}{\bf P}^{-1}_{K|K}\mathbf{D}_0^T$. So we just need to maximize ${\Sigma}_{eK}$ in order to maximize the determinant of ${\bf P}_{K|K}+{\bf A}_K$. This is equivalent to maximizing the trace of ${\bf P}_{K|K}+{\bf A}_K$ as discussed in Section \ref{sec:max_trace}.

\subsubsection{Attack Strategy For A Single Position And Velocity Sensor}
We assume that the adversary knows the system model and the prior information ${\bf P}_{0|0}$ at time zero, so that he/she can calculate the offline Kalman filter gain matrix ${\bf W}_k$ recursively. The best attack strategy is the solution to the following optimization problem.
	\beqa 
	\label{opt:single_vector}
		&&\max_{{\bf \Sigma}_{K}}  \left| {\bf {P}}_{K|K}+\mathbf{W}_{K}{\bf \Sigma}_{K}\mathbf{W}^{T}_{K} \right|\nn\\ 
		&& s.t. \;\;\; \sigma_{b_p}^2+T^2\sigma_{b_v}^2 = a^2\\
		&&\;\;\;\;\;\ -1\leq \rho_{b_p,b_v} \leq 1\nn\\
		&&\;\;\;\;\;\ \sigma_{b_p},\sigma_{b_v} > 0 \nn
	\eeqa  
where $ \mathbf{W}_{K}{\bf \Sigma}_{K}\mathbf{W}^{T}_{K}={\bf A}_{K}$, and 
\beq 
	\boldsymbol{\Sigma}_K=\left[\begin{array}{cc}
	\sigma_{b_p}^2& \rho_{b_p,b_v}\sigma_{b_p}\sigma_{b_v}\\
	\rho_{b_p,b_v}\sigma_{b_p}\sigma_{b_v}& \sigma_{b_v}^2\\
	\end{array}\right]
\eeq	
Using the properties of the determinant, we  get the formula as follows.
	\begin{eqnarray}
		&&|{\bf P}_{K|K}+\mathbf{W}_{K}{\bf \Sigma}_{K}\mathbf{W}^{T}_{K}|\nn\\
		&&=|{\bf P}_{K|K}||{\bf I}_n+{\bf \Sigma}_{K}\mathbf{W}^{T}_{K}{\bf P}^{-1}_{K|K}\mathbf{W}_{K}|
	\end{eqnarray}
Since ${\bf P}_{K|K}$ is independent of  ${\bf \Sigma}_{K}$, the optimization problem  can be further written as:  
\beqa 
\label{eq:det_pv_sen}
\max_{{\bf \Sigma}_{K}}	&& \left| {\bf I}_n+{\bf \Sigma}_{K}\mathbf{W}^{T}_{K}{\bf P}^{-1}_{K|K}\mathbf{W}_{K} \right|\nn\\ 
s.t.	&&  \;\;\; \sigma_{b_p}^2+T^2\sigma_{b_v}^2 = a^2\\
	&&\;\;\;\;\;\ -1\leq \rho_{b_p,b_v} \leq 1\nn\\
	&&\;\;\;\;\;\ \sigma_{b_p},\sigma_{b_v} > 0 \nn
\eeqa  

By defining 
\begin{eqnarray}
	\mathbf{W}^{T}_{K}{\bf P}^{-1}_{K|K}\mathbf{W}_{K} =\left[ \begin{array}{cc}
		m_{1} & m_{2}\\
		m_{2} & m_{3}\\
	\end{array}\right ]
\end{eqnarray}
and after simplifying (\ref{eq:det_pv_sen}), the objective function becomes 
\begin{eqnarray}
\label{eq:equi_1_sen_det}
	&& \left| {\bf I}_n+{\bf \Sigma}_{K}\mathbf{W}^{T}_{K}{\bf P}^{-1}_{K|K}\mathbf{W}_{K} \right|\nn\\
	&&=1+(1-\rho_{b_p,b_v}^2)\sigma_{b_p}^2\sigma_{b_v}^2(m_{1}m_{3}-m_{2}^2)\\
	&&+\sigma_{b_p}^2 m_{1}+\sigma_{b_v}^2 m_{3}+2\rho_{b_p,b_v}\sigma_{b_p}\sigma_{b_v}m_{2}\nn
\end{eqnarray} 
The optimal solution to the problem will be the best strategy to attack the system.

We denote ${\bf \Sigma}_{K}={\bf R}^T{\bf R}$ and since ${\bf \Sigma}_{K}$ is invertible, we have
\beqa 
	&& \left| {\bf I}_n+{\bf \Sigma}_{K}\mathbf{W}^{T}_{K}{\bf P}^{-1}_{K|K}\mathbf{W}_{K} \right|\nn\\
	=&& \left| {\bf I}_n+{\bf R}^T{\bf R}\mathbf{W}^{T}_{K}{\bf P}^{-1}_{K|K}\mathbf{W}_{K} \right|\\
	=&& \left| {\bf I}_n+{\bf R}\mathbf{W}^{T}_{K}{\bf P}^{-1}_{K|K}\mathbf{W}_{K}{\bf R}^T \right|\nn
\eeqa

In order to obtain the optimal solution, two useful lemmas \cite{Bo&Jun:TheMIMOradar2010} are introduced as follows,

\begin{lemma}
	Suppose ${\bf A}$ and ${\bf B}$ are $n\times n$ positive semidefinite matrices with eigendecomposition ${\bf A}={\bf \Psi}_{\bf A} {\bf \Sigma}_{\bf A} {\bf \Psi}^{T}_{\bf A}$ and ${{\bf B}={\bf \Psi}_{\bf B}{\bf \Sigma}_{\bf B}{\bf \Psi}^{T}_{\bf B}}$, the eigenvalues of ${\bf A}$ and ${\bf B}$ satisfy that $\alpha_{1}\ge\alpha_{2}\ge \cdots \ge \alpha_{n}$ and $\beta_{1}\ge\beta_{2}\ge \cdots \ge\beta_{n}$, then
	\begin{equation}
		\Pi_{i=1}^{n}(\alpha_{i}+\beta_{i})\le\det({\bf A+B})\le \Pi_{i=1}^{n}(\alpha_{i}+\beta_{n+1-i})
	\end{equation} 
	where the upper bound is achieved if and only if ${\bf \Psi_{A}=\Psi_{B}\Theta}$, the lower bound is achieved if and only if ${\bf \Psi_{A}=\Psi_{B}}$, and ${\bf \Theta}$ is the matrix defined below,
	\begin{equation}
			\left[\begin{array}{cccc}
				0& 0&\cdots& 1\\
				0& \cdots & 1& 0\\
				\vdots& \vdots& \vdots & \vdots\\
				1& 0&\cdots& 0	\end{array}	\right ]
	\end{equation}
\end{lemma}
Readers are referred to \cite{Bo&Jun:TheMIMOradar2010} for the proof of Lemma 1. The optimal solution to find the upper bound is the best strategy to attack the system with the most effect on Kalman filter system and the lower bound is the least attack effect the adversary can get.  

\begin{lemma}
	Given a $n \times n$ matrix ${\bf V}_{1}$ and a $n \times n$ positive semidefinite matrix ${\bf \Xi}_{1}$ with ${\bf V}_{1}{\bf \Xi}_{1}{\bf V}^{T}_{1}$ being a diagonal matrix with diagonal elements in increasing order, it is always possible to find another $n \times n$ matrix $\bar{{\bf V}}_{1}$ such that $\bar{{\bf V}}_{1}{\bf \Xi}_{1}\bar{{\bf V}}_{1}^T=\beta {\bf V}_{1}{\bf \Xi}_{1}{\bf V}^{T}_{1}$ with $Tr({\bf V}_{1}{\bf V}_{1}^T)=Tr(\bar{{\bf V}}_{1}\bar{{\bf V}}_{1}^T)$ where $\beta \ge 1$. $\bar{{\bf V}}_{1}$ can be written as $\Sigma_{{\bf \Xi}}{\bf \Psi}_{1}^T$, where ${\bf \Psi}_{1}$ is the unitary matrix whose columns are the eigenvectors corresponding to the eigenvalues of ${\bf \Xi}_{1}$ in increasing order, and $\bf \Sigma_{\Xi}$ is a diagonal matrix.
\end{lemma}

 By combining the two lemmas together, we can get the final optimal solution to the optimization problem above. It is obvious that ${\bf I}_n $ and ${\bf R}\mathbf{W}^{T}_{K}{\bf P}^{-1}_{K|K}\mathbf{W}_{K}{\bf R}^T$ are both positive semidefinite matrices, and their eigendecomposition can be written as follows,
\begin{eqnarray}
	&&{\bf I}_n= {\bf \Psi}_{1}{{\bf \Sigma}_{1}}{\bf \Psi}^T_{1} \nn \\
	&&{\bf R}\mathbf{W}^{T}_{K}{\bf P}^{-1}_{K|K}\mathbf{W}_{K}{\bf R}^T={\bf \Psi}_{2}{\bf \Sigma}_{2}{\bf \Psi}_{2}^T
\end{eqnarray}
with identity matrix ${\bf{\Sigma}}_{1}=diag([\sigma_{1,1},\cdots,\sigma_{1,n}])$ and ${\bf \Sigma}_{2}=diag([\sigma_{2,1},\cdots,\sigma_{2,n}])$, where $\sigma_{2,i}, i \in \{1,\cdots,n\}$ is the diagonal element of the matrix ${\bf \Sigma}_2$. Based on Lemma 1, we can get,
\begin{equation}
	\left| {\bf I}_n +{\bf R}\mathbf{W}^{T}_{K}{\bf P}^{-1}_{K|K}\mathbf{W}_{K}{\bf R}^T \right|\le \Pi_{i=1}^{n}(\sigma_{2,i}+1)
\end{equation}
where ${{\bf \Psi}_{1}={\bf \Psi}_{2}{\bf \Theta}}$.
\begin{eqnarray}
	&&|{\bf I}_n +{\bf R}\mathbf{W}^{T}_{K}{\bf P}^{-1}_{K|K}\mathbf{W}_{K}{\bf R}^T|\nn\\
	&&=|{\bf \Psi}_{1}^T||{\bf I}_n +{\bf R}\mathbf{W}^{T}_{K}{\bf P}^{-1}_{K|K}\mathbf{W}_{K}{\bf R}^T||{\bf \Psi}_{1}|\\
	&&=|{\bf I}_{n}+{\bf \Psi}^T_{1}{\bf R}\mathbf{W}^{T}_{K}{\bf P}^{-1}_{K|K}\mathbf{W}_{K}{\bf R}^T{\bf \Psi}_{1}|\nn
\end{eqnarray}
Set ${\bf R}_{1}={\bf \Psi}_{1}^T{\bf R}$ and ${\bf \Sigma}_{3}={\bf \Theta}{\bf \Sigma}_{2}{\bf \Theta}^T$ with the eigenvalues of increasing order and $Tr({\bf R}{\bf R}^T)=Tr({\bf R}_{1}{\bf R}_{1}^T)$. So the optimization problem can be written as below,
\begin{eqnarray}
	\max &&|{\bf I}_n +{\bf R}_{1}\mathbf{W}^{T}_{K}{\bf P}^{-1}_{K|K}\mathbf{W}_{K}{\bf R}_{1}^T|\nn\\
	\string s.t. && Tr({\bf R}_{1}{\bf R}_{1}^T) \le a^2\\
	&& {\bf R}_{1}\mathbf{W}^{T}_{K}{\bf P}^{-1}_{K|K}\mathbf{W}_{K}{\bf R}_{1}^T={\bf\Sigma_{3}}\nn
\end{eqnarray}
Setting $\mathbf{W}^{T}_{K}{\bf P}^{-1}_{K|K}\mathbf{W}_{K}=\tilde{{\bf \Xi}}$, we have $ {\bf R}_{1}\tilde{{\bf \Xi}}{\bf R}_{1}^T={\bf \Sigma}_{3}$. Based on Lemma 2, we can surely find a matrix $\bar{\bf R}$ such that ${\bf \bar{ R}}_{1}\tilde{{\bf \Xi}}{\bf \bar{R}}_{1}^T=\beta{\bf R}_{1}\tilde{{\bf \Xi}}{\bf R}_{1}^T$, with $\beta \ge 1$. Note that det($\cdot$) is a monotonic increasing function of the positive semidefinite matrix. So 
\begin{eqnarray}
	|{\bf I}_n +{\bf R}_{1}\tilde{{\bf \Xi}}{\bf R}_{1}^T|\le|{\bf I}_n+{\bf \bar{ R}}_{1}\tilde{{\bf \Xi}}{\bf \bar{R}}_{1}^T|
\end{eqnarray}
So the optimal solution $\bar{\bf R}$ should be in the form of $\bar{\bf V}$. The eigendecompostion of $\tilde{\bf \Xi}$ is as follows,
\begin{equation}
	\tilde{{\bf \Xi} }={\bf V}_{\Xi}{\bf \Sigma}_{\Xi}{\bf V}_{\Xi}^T
\end{equation}
where ${\bf \Sigma}_{\Xi}=diag([\sigma_{\xi,1},\sigma_{\xi,2},\cdots,\sigma_{\xi,n}])$ in increasing order. ${\bf V}_{\Xi}$ is a unitary matrix whose column vectors corresponds to the eigenvalues of $\tilde{{\bf \Xi} }$. The problem can be written as 
\begin{eqnarray}
	\max_{\sigma^2_{b,i}} &&\sum_{i=1}^{n}log(\sigma^2_{b,i}\sigma_{\xi,i}+1)\\
	s.t. && \sum_{i=1}^{n}(\sigma^2_{b,i}) \le a^2\nn
\end{eqnarray}
The objective function above is a concave and increasing function. The optimal solution is achieved through Lagrangian multipliers yielding the water-filling strategy,
\begin{equation}
	\sigma^2_{b,i}=\left(\frac{1}{\lambda}-\frac{1}{\sigma_{\xi,i}}\right)^+
\end{equation}
where the value of $\lambda$ can be obtained by solving
\begin{equation}
	\sum_{i=1}^{n}\left(\frac{1}{\lambda}-\frac{1}{\sigma_{\xi,i}}\right)^+=a^2
\end{equation}
The solution is 
\begin{equation}
	{\bf R}^{opt}={\bf \Psi}_{1}[\bf{\Sigma}_b^{1/2}]^T{\bf V}^T_{\Xi}
\end{equation}
Finally, the optimal solution of (\ref{eq:det_pv_sen}) is,
\begin{eqnarray}
	{\bf \Sigma}_{K}=	{\bf V}_{\bf \Xi}{\bf \Sigma}_b{\bf V}^T_{\bf \Xi}
\end{eqnarray}

\subsubsection{Attack Strategy For Multiple Position and Velocity Sensors}
For a system with multiple sensors, the best strategy to allocate the bias noise power and set the correlation coefficients among the bias noises at different sensors is also investigated. Let us denote  the number of sensors  as  $M$, and the measurement matrix as ${\bf H}=[{\bf I}_2, \cdots,  {\bf I}_2]^{T}$. The measurement covariance matrix for the $i$th sensor is assumed to be 
\beq
	{\bf R}_{i}=\left
	[\begin{array}{ccc}
		\sigma^{2}_{p_i}& 0\\
		0& \sigma^{2}_{v_i}\\
	\end{array}\right]
\eeq

Now, according to (\ref{eq:rnew}), we have 
\beq
	\begin{split}
		{\bf R}_{e}&=\left(\sum_{i=1}^{M}{\bf R}^{-1}_{i}\right)^{-1}\\
		&=\left[
		\begin{array}{cc}
			\left(\sum_{i=1}^{M}\sigma^{-2}_{p_i}\right)^{-1}& 0\\
			0& \left(\sum_{i=1}^{M}\sigma^{-2}_{v_i}\right)^{-1}\\
		\end{array}\right]
	\end{split}
\eeq
According to (\ref{eq:ze}), we define 
\beq
	\begin{split}
		{\bf C}_{i}&={\bf R}_{e} {\bf R}^{-1}_{i}\\
		&=\left[
		\begin{array}{ccc}
			\frac{\sigma^{-2}_{p_i}}{\sum_{j=1}^{M}\sigma^{-2}_{p_j}}& 0\\
			0 &\frac{\sigma^{-2}_{v_i}}{\sum_{j=1}^{M}\sigma^{-2}_{v_j}}\\
		\end{array}\right]
	\end{split}
\eeq
as the weighting matrix for the $i$th sensor's observation ${\bf z}_i$.

The equivalent injected bias noise is therefore
\beq 
	{\bf b}_{eK}=\sum_{i=1}^{M} {\bf C}_i {\bf b}_{K_i}
\eeq
and  the covariance matrix of the equivalent bias vector is
\beq
	\label{eq:Sigma_eq_two_1}
	{\bf \Sigma}_{eK}=\sum_{i=1}^{M} \sum_{j=1}^{M} {\bf C}_i  E(b_{i}b^{T}_{j}) {\bf C}_j^T
\eeq
Now the optimization problem can be formulated as follows.
\beqa
	\label{eq:max_det_vec_multisensor_case}
	&&\max_{{\bf \Sigma}_{eK}}  \left| {\bf P}_{K|K}+\mathbf{W}_{eK}{\bf \Sigma}_{eK}\mathbf{W}^{T}_{eK} \right|\\ 
	&& s.t. \;\;\; \sum_{i=1}^{N}\sigma_{b_{p_i}}^2+T^2\sum_{j=1}^{N}\sigma_{b_{v_i}}^2= a^2,\nn\\
	&&\;\;\;\;\;\ -1\leq \rho_{b_{p_i},b_{v_j}} \leq 1, \nn\\
	&&\;\;\;\;\;\ -1\leq \rho_{b_{v_i},b_{v_j}} \leq 1, \nn\\
	&&\;\;\;\;\;\ -1\leq \rho_{b_{p_i},b_{p_j}} \leq 1, \nn\\
	&&\;\;\;\;\;\ \sigma_{b_{p_i}},\sigma_{b_{v_i}}\geq 0, \;\;\;\forall i,j \in \{1,M\}\nn
\eeqa 
where ${\bf W}_{eK}$ is Kalman filter gain calculated using ${\bf H}_e$ and ${\bf R}_e$. The optimal solution of  (\ref{eq:max_det_vec_multisensor_case}) can  be obtained numerically as shown later in the paper. 
\section{Numerical Results}
\label{sec:numerical}
Some numerical results are presented in this section to illustrate the theoretical results.  
\subsection{System with Position Sensors}
The parameters used in the target tracking example   are provided below. The system sampling interval  is $T=1$.  The adversary injects bias information to two sensors with $\sigma_{w_1}^{2} =3$ and $\sigma_{w_2}^{2} =4$,  respectively. The variance of the system process noise is $\sigma_{v}^{2} =0.25$. The biases $b_{i}$s are zero-mean Gaussian random variables with variances $\sigma^{2}_{b_i}$s. For the power constraint we discussed earlier, we set the sum of $\sigma^{2}_{b_i}$ to be 3000. 

\begin{figure}[htb]
	\centering
	{\includegraphics[width=3.2in]{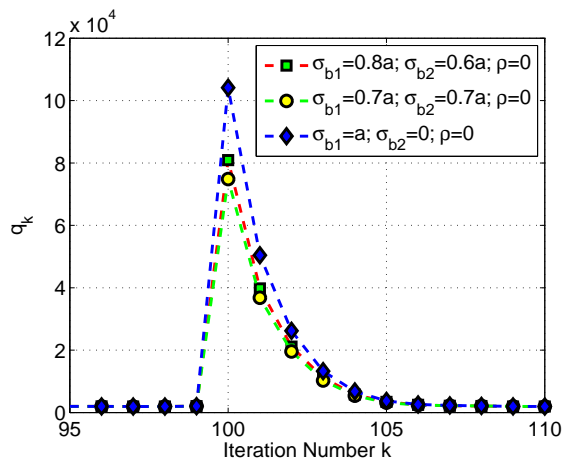}}
	\caption{The normalized MSE when independent biases are used. $\sigma^2_{b_1}+\sigma^2_{b_2}=a^2$ for each case.}
	\label{fig:Single Injection1}
\end{figure}

The effect of the bias injection on Kalman filter is measured by a Chi-squared test. More specifically, we use the sum of the normalized MSE over $N_{m}$ Monte-Carlo runs 
\begin{eqnarray}
q_k= \sum^{N_{m}}_{j=1} \left[ \hat{ \mathbf{x}}'^j_{k|k} - \mathbf{x}^j_{k} \right]^{T} \mathbf{P}_{k|k}^{-1}\left[ \hat{ \mathbf{x}}'^j_{k|k} - \mathbf{x}^j_{k} \right]
\label{eq:}
\end{eqnarray}
\noindent 
where at time $k$, $\mathbf{P}_{k|k}$ is the nominal state covariance matrix calculated by Kalman filter, $\hat{ \mathbf{x}}'^j_{k|k}$ is the state estimate, and $\mathbf{x}^j_{k}$ is the true state, during the $j$th Monte-Carlo run. First, if the random biases injected to different sensors are independent, 
we should allocate all the bias power to the sensor with the smallest measurement noise variance. This is clearly true as demonstrated in Fig. \ref{fig:Single Injection1}, where  allocating all the power to sensor 1 causes the maximum estimation MSE.
\begin{figure}[htb]
	\centering
	\caption{The normalized MSE for dependent biases. $\sigma^2_{b_1}+\sigma^2_{b_2}=a^2$ for each case.}
	\label{fig:Single_Injection3}
\end{figure} 
In Fig. \ref{fig:Single_Injection3}, three dependent-noise attack strategies are compared, including  the optimal one according to (\ref{op_solu_2}), allocating the power equally among the sensors, and allocating all the power to the sensor with smallest measurement error variance.  It is clear that the optimal solution has the largest impact on the estimation performance, and it outperforms the best independent-noise attack strategy significantly.
 
\subsection{Systems with Position and Velocity Sensors}
\begin{figure}[htb]
	\centering
	{\includegraphics[width=2.88in]{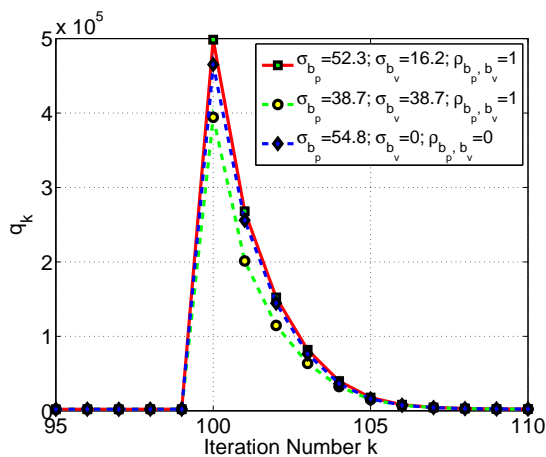}}
	\caption{The normalized MSE for a system with a single sensors. $\sigma^2_{p_1}+T^2\sigma^2_{v_1}=a^2$ for each case.}
	\label{fig:vector_injection}
\end{figure}

We now consider the case where the adversary attacks Kalman filtering system with a vector sensor observation containing both position and velocity measurements. We first consider a single-sensor system, and the sensor has a position measurement variance of 3 and a velocity measurement variance of 4. We set the sum of $\sigma_{b_{p_1}}^2$ and $T^2\sigma_{b_{v_1}}^2$ to be 3000. In this particular case, $w_{11}w_{12}+w_{21}w_{22}>0$, so the optimal choice is  $\rho_{b_p,b_v}=1$.  Based on Theorem \ref{the:single_vector},  the best strategy is to set $\sigma_{b_p}=52.3$ and $\sigma_{b_v}=16.2$. It is clear from  Fig. \ref{fig:vector_injection} that the strategy provided in Theorem \ref{the:single_vector} maximizes the MSE of Kalman filter system by injecting vector bias information.

\begin{figure}[htb]
	\centering
	{\includegraphics[width=4.2in]{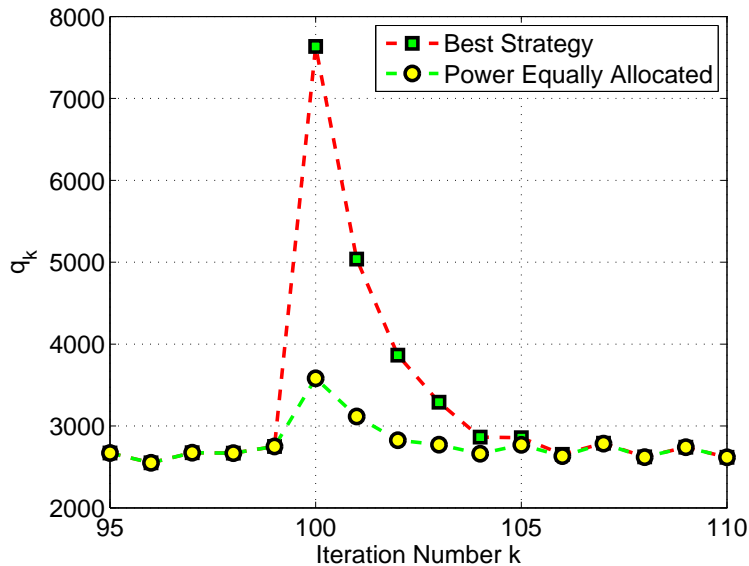}}
	\caption{The normalized MSE for a system with two sensors. $\sigma_{p_1}^2+\sigma_{p_2}^2+T^2\sigma_{v_1}^2+T^2\sigma_{v_2}^2 = a^2$ for each case.}
	\label{fig:2vector_injection}
\end{figure}

Next we consider a system with two sensors. The first sensor is the same as the one described above, and the second one is with position measurement variance 4 and velocity measurement variance 5. In this particular case, again we have $w_{11}w_{12}+w_{21}w_{22}>0$, so all the $\rho$s in $s_1$, $s_2$, and $s_3$ should be set as 1.  We first use a systematic grid search to find an approximate globally optimal solution and then we use the FMINCON function in Matlab, a local search algorithm, to refine this approximate globally optimal solution. The optimal solution we have obtained is $\sigma_{b_{p_1}}^2=1826,\sigma_{b_{p_2}}^2=1023,\sigma_{b_{v_1}}^2=81,\sigma_{b_{v_2}}^2=68$. For comparison purpose, we also implement an attack strategy that allocate power equally among the observation components and among the two sensors, which is $\sigma_{b_{p_1}}^2=\sigma_{b_{p_2}}^2=\sigma_{b_{v_1}}^2=\sigma_{b_{v_2}}^2=750$. 
The simulation result is shown in  Fig. \ref{fig:2vector_injection}.  As we can see, the optimal attack strategy has a much greater impact than the one that allocates power equally. Based on the optimal solution, we can find that allocating more power to the measurement with lower variance will have a greater effect on Kalman filter system. 

\subsection{Determinant case}
\begin{figure}[htb]
		\centering
		{\includegraphics[width=3.2in]{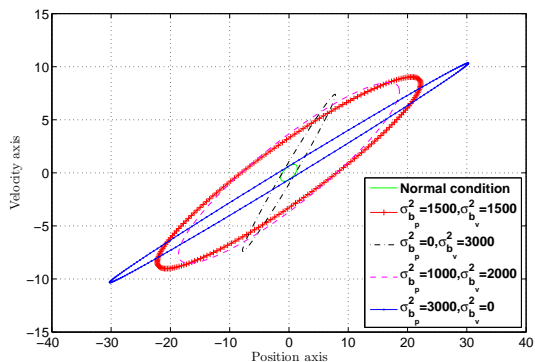}}
		\caption{Error ellipsis for  different power allocation strategies}
		\label{fig:Vec_case_Single_Power}
	\end{figure}

\begin{figure}[htb]
	\centering
	\includegraphics[width=3.2 in]{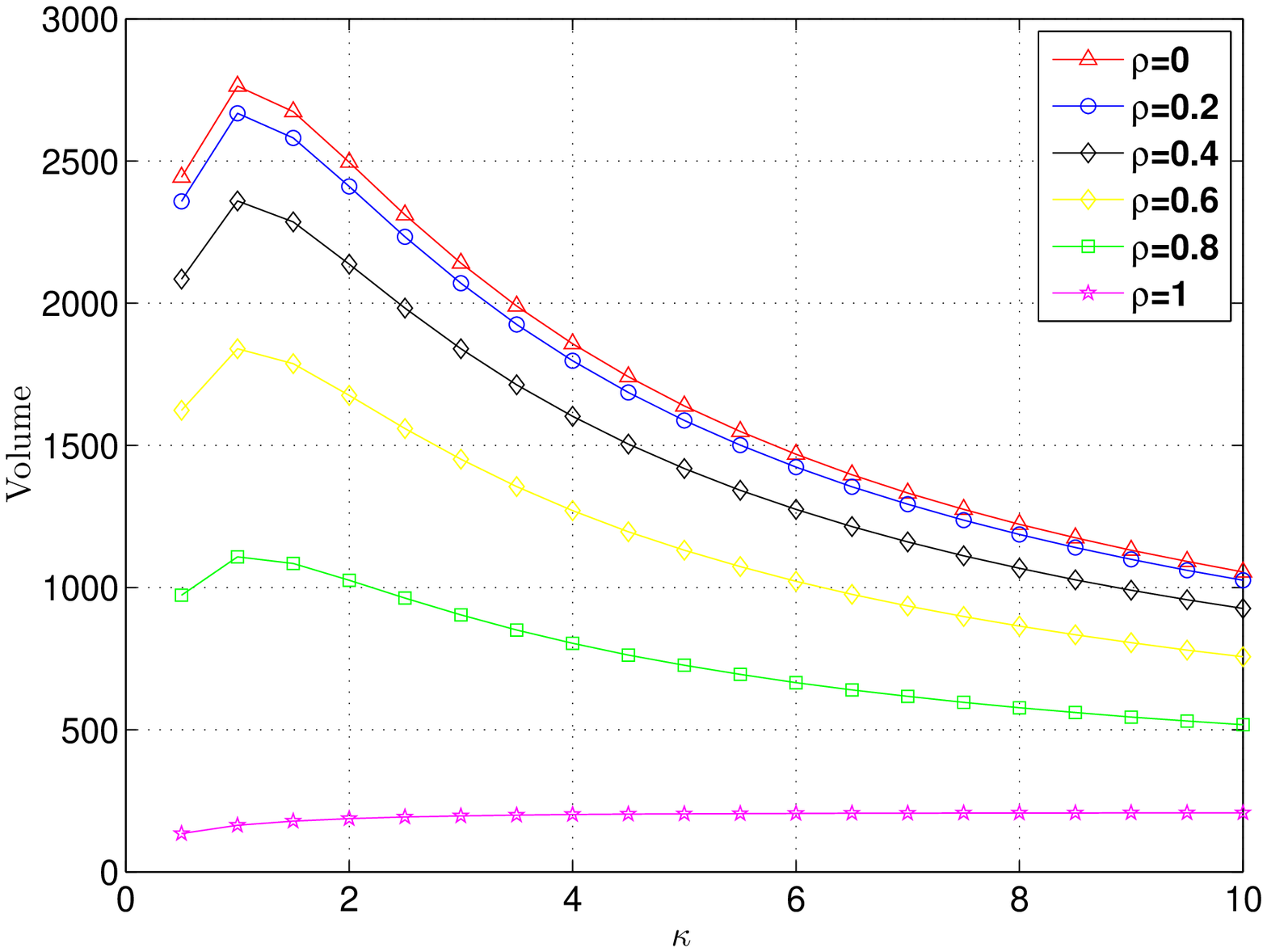}
	\caption{Error ellipse volume}
	\label{fig:Volume_positive_rho}
\end{figure}

\begin{figure}[htb]
	\centering
	\includegraphics[width=3.2 in]{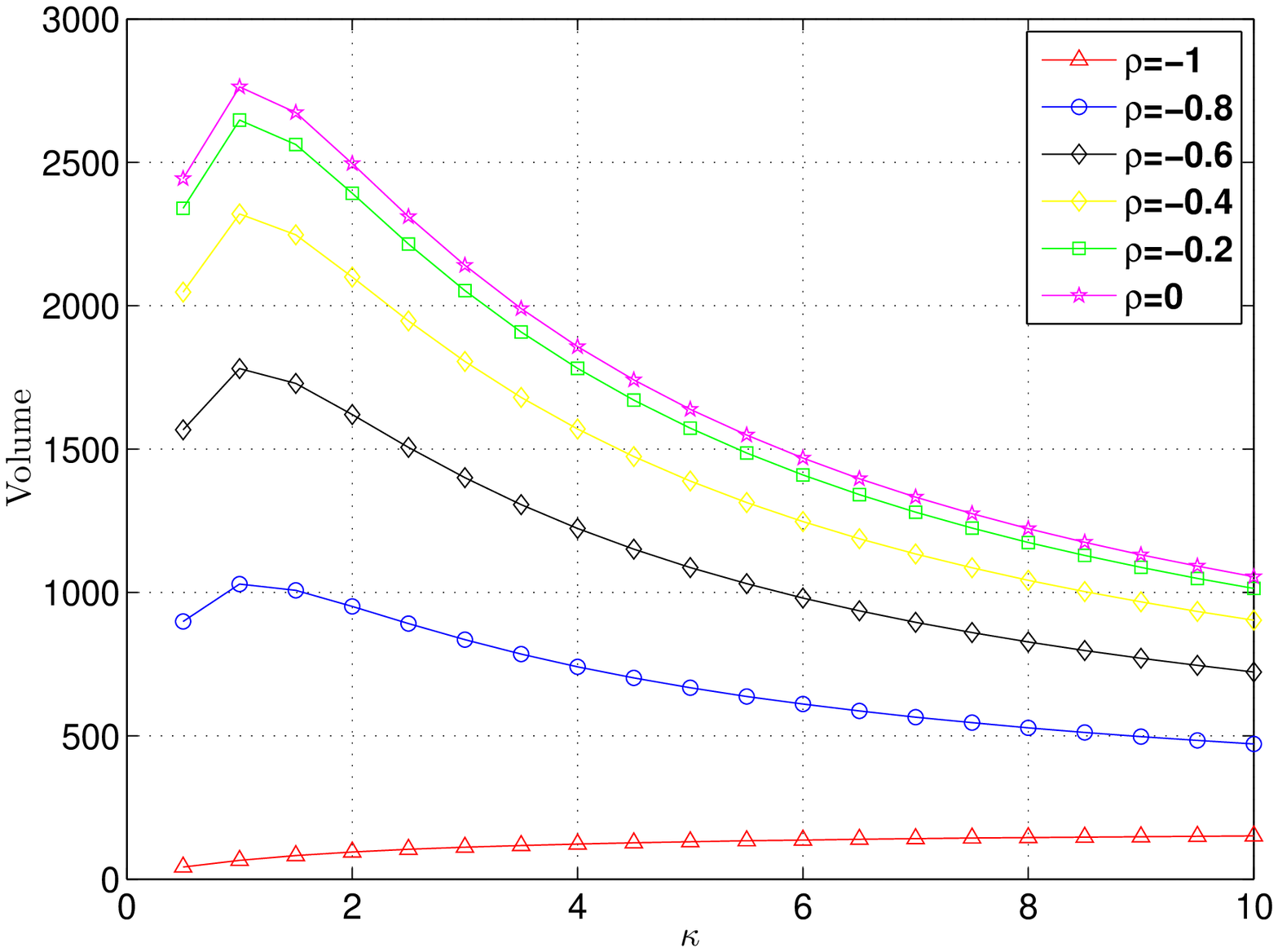}
	\caption{Error ellipse volume}
	\label{fig:Volume_negative_rho}
\end{figure}

\begin{figure}[htb]
	\centering
	{\includegraphics[width=3.2in]{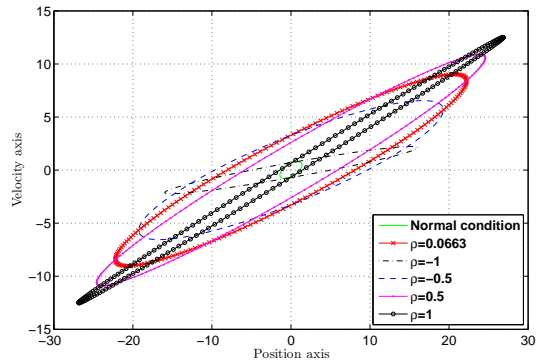}}
	\caption{Error ellipsis for different $\rho$s}
	\label{fig:Vec_case_Single_Rho}
\end{figure}

Numerical results are presented in this section to illustrate the effectiveness of the proposed attack strategies. Assuming  that the injected bias noise ${\bf  b}_{k}$ is zero-mean and Gaussian distributed, we can show that the posterior probability density function (PDF) of the target state conditioned on the past observations and the current corrupted observation is 
\begin{equation}
	p({\bf x}_{K}|{\bf z}_{1:K-1}, {\bf z}'_K)= {\cal N} (\hat{\bf x}_{K|K}, {\bf P}_{K|K}+{\bf A}_K )
\end{equation}
where $\hat{\bf x}_{K|K}$ is the updated state estimate calculated by Kalman filter, which is unaware of the presence of the injected false information. 
Then the target state ${\bf x}_K$ will be in the following confidence region (or error ellipse) 
\beqa
	\label{eq:volume}
	\left\lbrace {\bf x}: ({\bf x}-\hat{\bf x}_{K|K})^T ({\bf P}_{K|K}+{\bf A}_K)^{-1} ({\bf x}-\hat{\bf x}_{K|K})\le\gamma\right\rbrace
\eeqa
with probability determined by the threshold $\gamma$ \cite{bar-shalom&etal:book11}. The volume of the confidence region defined by (\ref{eq:volume}) corresponding to the threshold $\gamma$ is
\begin{equation}
	V(K)=c_{n_x}|\gamma ({\bf P}_{K|K}+{\bf A}_K)|^{1/2}
\end{equation}
where $n_x$ is the dimension of the target state  ${\bf x}$, 
\begin{eqnarray}
	c_n=\frac{\pi^{n/2}}{\Gamma(n/2+1)}
\end{eqnarray}
and $\Gamma(\cdot)$ is the gamma function.
 First, let us consider  a single-sensor case, where the sensor has a position measurement with noise variance of $3$, which is independent of the velocity measurement with noise variance of $4$. We set the bias noise power constraint as  $\sigma_{b_{p}}^2+ T^2\sigma_{b_{v}}^2=3000$. We solve the optimization problem formulated in Section \ref{sec:single-sensor} numerically, and the optimal solution to (\ref{opt:single_vector}) is  $\sigma^{2}_{b_p}=1500, \sigma^{2}_{b_v}=1500, \rho_{b_{p,v}}=0.063$. In Fig. \ref{fig:Vec_case_Single_Power}, error ellipsis for different attack strategies are plotted. For all the different attack strategies, we set $\rho_{b_{p,v}}=0.063$. 
As we can see, under normal condition without false information injection,  the error ellipse has the smallest area, while the optimal attack strategy leads to an error ellipse with the largest area. In Figs. \ref{fig:Volume_positive_rho} and \ref{fig:Volume_negative_rho}, the volume (area) of the error ellipse is provided as a function of $\rho_{b_{p,v}}$  and the ratio  $\kappa=\frac{\sigma_{b_p}}{\sigma_{b_v}T}$.  We can see that when the $\kappa=\frac{\sigma_{b_p}}{\sigma_{b_v}T}=1$, the area of the ellipse is maximized. Also from Figs. \ref{fig:Volume_positive_rho} and \ref{fig:Volume_negative_rho}, it is clear that the area of ellipse increases as the absolute value of $\rho $ decreases. In Fig. \ref{fig:Vec_case_Single_Rho}, the trend of the error ellipsis as the $\rho $ changes from $-1$ to $+1$ is illustrated.	


\begin{figure}[htb]
	\centering
	{\includegraphics[width=3.2in]{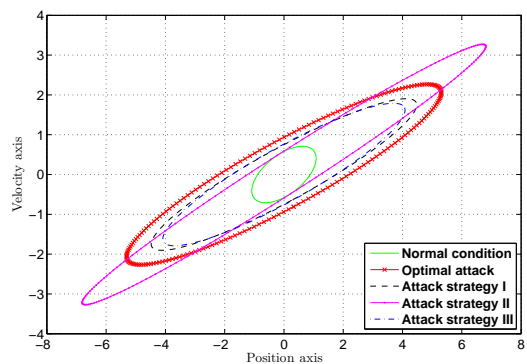}}
	\caption{Error ellipsis for different power allocation strategies}
	\label{fig:vec_case_twosensor}
\end{figure}

In this particular case, since $\sigma_{b_{p}}^2+ T^2\sigma_{b_{v}}^2=3000$, $\boldsymbol{\Sigma}_K$ is large and in (\ref{opt:single_vector}) the second term ($\mathbf{W}_{K}{\bf \Sigma}_{K}\mathbf{W}^{T}_{K}$) dominates.  Therefore, in (\ref{eq:equi_1_sen_det})  the identity matrix in the objective function is relatively small  comparing to the second item, and approximately we have
 \begin{eqnarray}
 &&\left| {\bf I}_n+{\bf \Sigma}_{K}\mathbf{W}^{T}_{K}{\bf P}^{-1}_{K|K}\mathbf{W}_{K} \right|\nn \\
  &&\approx \left| {\bf \Sigma}_{K}\right|\left| \mathbf{W}^{T}_{K}{\bf P}^{-1}_{K|K}\mathbf{W}_{K}  \right|
 \end{eqnarray} 
The second term in the second line of the above equation is a constant. Hence, in order to get the maximum determinant, we should set $\sigma_{b_p}^2=\sigma_{b_v}^2 T^2$ and $ \rho_{b_p,b_v}=0$. This is almost the same solution as we have obtained numerically.
Next we consider a system with two sensors. The first sensor is the same as the one described above, and the second one is with position measurement variance $4$ and velocity measurement variance $5$. To solve the optimization problem formulated in (\ref{eq:max_det_vec_multisensor_case}), we first use a systematic grid search to find an approximate globally optimal solution and then we use the FMINCON function in Matlab, a local search algorithm, to refine this approximate globally optimal solution. The optimal solution we have obtained is $\sigma_{b_{p_1}}^2=1100,\;\sigma_{b_{p_2}}^2=600,\;\sigma_{b_{v_1}}^2=750,\;\sigma_{b_{v_2}}^2=550$,\; $\rho_{b_{p_1,p_2}}=0.99,\;\rho_{b_{p_1,v_1}}=-0.83,\;\rho_{b_{p_1,v_2}}=0.75,\; \rho_{b_{v_1,p_2}}=0.89,\;\rho_{b_{p_2,v_2}}=-0.23,\;\rho_{b_{v_1,v_2}}=0.95$.  For comparison purpose, we introduce three sub-optimal attack strategies: Strategy I with all the $\rho$s being $0$s, and $\sigma_{b_{p_1}}^2=1100,\sigma_{b_{p_2}}^2=600,\sigma_{b_{v_1}}^2=750,\sigma_{b_{v_2}}^2=550$; Strategy II with all the $\rho$s being $1$s, and $\sigma_{b_{p_1}}^2=1100,\sigma_{b_{p_2}}^2=600,\sigma_{b_{v_1}}^2=750,\sigma_{b_{v_2}}^2=550$; and Strategy II with the $\rho$s being the same as those for the optimal strategy, and $\sigma_{b_{p_1}}^2=\sigma_{b_{p_2}}^2=\sigma_{b_{v_1}}^2=\sigma_{b_{v_2}}^2=750$. The numerical results are shown in  Fig. \ref{fig:vec_case_twosensor}.  As we can see, the optimal attack strategy has a greater impact than  those sub-optimal attack strategies, resulting in the largest error ellipse.
\section{Conclusions}
In this paper, we derived the EMSE due to the injected random biases for a Kalman filter in a linear dynamic system. This allows us to find how to allocate the bias power among multiple sensors in order to maximize the effect of the false information on Kalman filter from two perspectives: trace and determinant. A concrete example of multi-sensor target tracking system has been provided. In this example, we investigated both the case where the sensors provide position measurements and the case where they collect both position and velocity measurements. Further, many closed-form results have been provided for the optimal attack strategies. In the future, we will use game theory and hypothesis testing techniques to characterize the model in order to have a better understanding of the false information attacks and Kalman filter defense against such attacks.
\bibliographystyle{IEEEbib}
\bibliography{Book,Journal,Conf,Misc}

\end{document}